\documentclass[final,5p,times,twocolumn]{elsarticle}

\usepackage{lipsum}
\usepackage{ulem}
\usepackage{xcolor}
\usepackage[colorlinks=true]{hyperref}
\usepackage{soul}
\usepackage{booktabs}
\usepackage{mathrsfs}
\usepackage{amssymb}

\usepackage{amsmath,amssymb,color,epsfig}
\allowdisplaybreaks[4]

\newcommand{\be}{\begin{equation}}
\newcommand{\ee}{\end{equation}}
\newcommand{\bea}{\setlength\arraycolsep{2pt} \begin{eqnarray}}
\newcommand{\eea}{\end{eqnarray}}
\newcommand{\nn}{\nonumber}

\def\0{{\sst{(0)}}}
\def\1{{\sst{(1)}}}
\def\2{{\sst{(2)}}}
\def\3{{\sst{(3)}}}
\def\4{{\sst{(4)}}}
\def\5{{\sst{(5)}}}
\def\6{{\sst{(6)}}}
\def\7{{\sst{(7)}}}
\def\8{{\sst{(8)}}}
\def\sst#1{{\scriptscriptstyle #1}}

\journal{Physics Letters B}

\begin{document}

\begin{frontmatter}



\title{Revisiting black holes and their thermodynamics in Einstein-Kalb-Ramond gravity}


\author[inst1]{Zhong-Xi Yu}
\ead{zhongxiyu@yau.edu.cn}

\author[inst2]{Hong-Da Lyu}
\ead{hongdalyu@sdu.edu.cn}

\author[inst1]{Mandula Huhe}

\author[inst3]{Shoulong Li\corref{cor3}}
\cortext[cor3]{Corresponding author. Email: shoulongli@hunnu.edu.cn}

\affiliation[inst1]{organization={College of Physics and Electronic Information Engineering, Jining Normal University},
            addressline={}, 
            city={Wulanchabu},
            postcode={012000}, 
            state={Inner Mongolia},
            country={China}}

\affiliation[inst2]{organization={Key Laboratory of Particle Physics and Particle Irradiation (MOE),
Institute of Frontier and Interdisciplinary Science, Shandong University},
            addressline={}, 
            city={Qingdao},
            postcode={266237}, 
            state={Shandong},
            country={China}}

\affiliation[inst3]{organization={Department of Physics, Key Laboratory of Low Dimensional Quantum Structures and Quantum Control of Ministry of Education, and Institute of Interdisciplinary Studies, Hunan Normal University},
            addressline={}, 
            city={Changsha},
            postcode={410081}, 
            state={Hunan},
            country={China}}

\affiliation[inst4]{organization={Hunan Research Center of the Basic Discipline for Quantum Effects and Quantum Technologies, Hunan Normal University},
            addressline={}, 
            city={Changsha},
            postcode={410081}, 
            state={Hunan},
            country={China}}

\begin{abstract}
Einstein-Kalb-Ramond (EKR) gravity is an alternative theory in which a rank-two antisymmetric tensor field, the Kalb-Ramond field, is nonminimally coupled to gravity, potentially generating Lorentz-violating backgrounds. In this work, we revisit black hole solutions and thermodynamics in EKR gravity, addressing subtleties overlooked in previous studies. We obtain two distinct classes of exact static black hole solutions with general topological horizons in diverse dimensions, both with and without a cosmological constant, corresponding to different coupling sectors dictated by the field equations. We analyze their thermodynamic properties and, using the Wald formalism, compute the Noether mass and entropy, establishing the first law and clarifying the role of the Noether mass. Finally, we discuss the implications of this definition of mass for observational constraints in EKR gravity.

\end{abstract}




\end{frontmatter}




\section{Introduction}
\label{introduction}

Einstein-Kalb-Ramond (EKR) gravity~\cite{Altschul:2009ae}, featuring a rank-two antisymmetric Kalb-Ramond (KR) field nonminimally coupled to gravity~\cite{Kalb:1974yc}, has attracted increasing attention. With a suitable potential, the theory admits simple exact black hole solutions~\cite{Yang:2023wtu} in which certain components of the tensor field acquire nonvanishing vacuum values throughout spacetime. Although the Lagrangian itself remains generally covariant, the nonzero vacuum configuration of the KR field selects preferred tensorial background in the local spacetime frame, which persists even in the asymptotic region. This is commonly interpreted as a spontaneous breaking of local Lorentz symmetry~\cite{Altschul:2009ae}. The corresponding Lorentz-violating effects are characterized by dimensionless parameters constructed from the coupling constant and the asymptotic vacuum amplitude of the KR field. These black hole solutions and their associated thermodynamic quantities---particularly the gravitational mass---provide a practical framework for testing Lorentz symmetry breaking in the gravitational sector through black hole physics and astrophysical phenomena. Constraints on the Lorentz-violating parameter have already been obtained from Solar-System tests, among which the Shapiro time-delay measurement gives the most stringent bounds, restricting the relevant parameter to a narrow interval between $-10^{-13}$ and $10^{-14}$~\cite{Yang:2023wtu}. These developments have further stimulated interest in the theory. However, the nonminimal coupling between the gravitational and 2-form fields introduces subtleties that have not been fully addressed.

To discuss these subtleties more concretely, let us recall that the Lagrangian density of EKR gravity takes the form~\cite{Altschul:2009ae}
\be
L = R - 2\Lambda + \gamma_1 B^{\mu\lambda} B^\nu{}_\lambda R_{\mu\nu} +\gamma_2 B^2 R  -\frac{1}{12} H^2 -V(B^2) \,,  \label{gravity}
\ee
where $R$ and $R_{\mu\nu}$ denote the Ricci scalar and the Ricci tensor, respectively, and $\Lambda$ is the cosmological constant. The KR field $B_{\mu\nu}$ is nonminimally coupled to both the Ricci tensor and Ricci scalar through the coupling constants $\gamma_1$ and $\gamma_2$, while its field strength and potential are denoted by  $H_{\mu\nu\lambda} = 3\partial_{[\mu} B_{\nu\lambda]}  $ and $V$, respectively. 

A first subtlety concerns the treatment of the nonminimal coupling terms in the Lagrangian. In an early study~\cite{Lessa:2019bgi}, it was argued that, when the KR field acquires a nonzero value, the term $\gamma_2 B^2 R$ can be absorbed into the Einstein-Hilbert term via a suitable redefinition. While this captures part of the effect, it does not provide a general treatment of the coupled system. For $\gamma_1 = 0$, the curvature part of the Lagrangian reduces to $\Phi R$ and can be mapped to GR in vacuum via a conformal-like redefinition. By contrast, for $\gamma_1 \ne 0$, the gauge symmetry is broken and such a mapping fails. Moreover, even in the $\gamma_1 = 0$ case, the vacuum equivalence does not generally persist in the presence of matter, e.g., in compact stars.

A second subtlety arises from the fact that the equation of motion (EOM) of the KR field must be carefully verified. Since it generally involves curvature terms, it introduces additional constraints beyond the gravitational field equations, unlike minimally coupled matter fields in GR. However, in order to obtain exact solutions, earlier studies often adopted an approach in which certain preferred nonvanishing components---together with explicit functional forms---were assigned to the KR field. This does not ensure that the KR field EOM is satisfied automatically, and in some cases it was not explicitly verified, potentially leading to incomplete conclusions.

A third subtlety concerns black-hole thermodynamics, particularly the definition of the gravitational mass relevant for observations. Nonminimal couplings generally modify thermodynamic quantities, as also seen in scalar- and vector-coupled gravity~\cite{Feng:2015oea,Feng:2015wvb,An:2024fzf,Chen:2025ypx,Li:2025tcd}. Previous EKR studies often adopt standard GR definitions of mass and entropy~\cite{Yang:2023wtu}. A consistent treatment, however, requires a more general framework, e.g., the Wald formalism~\cite{Wald:1993nt,Iyer:1994ys}. An inappropriate identification of the physical gravitational mass may, in turn, lead to potentially unreliable constraints on the theoretical parameters inferred from astrophysical observations.

In this work, we revisit black hole solutions and thermodynamics in EKR gravity and carefully address these three subtleties. The paper is organized as follows. Sec.~\ref{flatsolution} presents the solutions without a cosmological constant. Sec.~\ref{Thermodynamics} analyzes their thermodynamics and obtains the Noether mass. Sec.~\ref{solar} discusses the implications of the Noether mass for observational constraints. Sec.~\ref{adssolution} extends the solutions and their thermodynamics to the case with a nonzero cosmological constant. Sec.~\ref{conclusion} concludes.

\section{$D$-dimensional Black hole solutions with $\Lambda=0$} \label{flatsolution}

The EOMs for $g_{\mu\nu}$ and $B_{\mu\nu}$ read
\begin{align}
&G_{\mu\nu}  + \gamma_1 \bigg[-\frac12 g_{\mu\nu} R_{\rho\sigma} B^{\rho}{}_\lambda B^{\sigma\lambda} +2 R_{\rho(\mu} B_{\nu)\sigma} B^{\rho\sigma} -R^{\rho\sigma} B_{\rho(\mu} B_{\nu)\sigma}  \nn\\
& +\frac12 g_{\mu\nu} \nabla_\rho\nabla_\sigma (B^\rho{}_\lambda B^{\sigma\lambda}) +\frac12 \nabla_\rho \nabla^\rho (B_{(\mu}{}^\lambda B_{\nu)\lambda}) - \nabla_\rho\nabla_{(\mu} (B_{\nu)\sigma} B^{\rho\sigma}) \bigg]  \nn\\
&  + \gamma_2 \bigg[ 2 R B_{(\mu}{}^\lambda B_{\nu)\lambda}  + g_{\mu\nu}\nabla_\lambda \nabla^\lambda X - \nabla_\mu\nabla_\nu X + G_{\mu\nu} X \bigg]  + \frac12 g_{\mu\nu} V \nn \\
& -\frac{1}{12} \bigg[3 H_{\mu\rho\sigma}H_\nu{}^{\rho\sigma} -\frac12 g_{\mu\nu} H^2 \bigg]  - 2 \frac{\partial V}{\partial X} B_{(\mu}{}^\lambda B_{\nu)\lambda} + \Lambda g_{\mu\nu} =0   \,, \label{eom1} \\
& 2 \gamma_1 R^{\lambda[\mu}B_\lambda{}^{\nu]} +2 \gamma_2 R B^{\mu\nu} + \frac12 \nabla_\lambda H^{\lambda\mu\nu} - 2 \frac{\partial V}{\partial X} B^{\mu\nu}  = 0 \,. \label{eom2} 
\end{align}
Here $X \equiv B^2 = B_{\mu\nu} B^{\mu\nu}$, $(\mu\nu)$ and $[\mu\nu]$ denote symmetrization and antisymmetrization. In this section, we set $\Lambda=0$. We consider static solutions with general horizon topology in $D=n+2$ $(n\ge2)$,
\begin{align}
&ds^2 = - h(r) dt^2 + f(r)^{-1} dr^2  + r^2 d\Omega_{n, k}^2  \,,  \label{metric}  \\
&\textup{with} \quad d\Omega_{n, k}^2 = \frac{du^2}{1-k u^2} +(1-k u^2) d\Omega_{n-1, k}^2  \,,  \label{topology}  
\end{align}
where $k=1,0,-1$ corresponds to spherical, planar, and hyperbolic geometries. Since the nonminimal couplings in EKR gravity can invalidate the Birkhoff theorem, a general ansatz~(\ref{metric}) without imposing $h=f$ a priori is useful for constructing solutions.
We assume the KR field develops a nonvanishing vacuum expectation value (VEV), $\langle B_{\mu\nu}\rangle=b_{\mu\nu}$.
Following Ref.~\cite{Lessa:2019bgi}, the antisymmetric two-form $b_{\mu\nu}$ 
can be decomposed, with respect to a timelike vector $v^\mu$, into 
pseudo-electric and pseudo-magnetic parts, $b_{\mu\nu}=\tilde E_{[\mu}v_{\nu]}+\epsilon_{\mu\nu\alpha\beta}v^\alpha \tilde B^\beta $, where $\tilde E_\mu v^\mu=0$ and $\tilde B_\mu v^\mu=0$. Here we choose the radial pseudo-electric configuration and set the pseudo-magnetic part to zero. This gives
\be
B_{\mu\nu} = b_{\mu\nu} = -\beta (r) dt \wedge dr \,, \label{bfield}
\ee
with potential $V \equiv V (X \pm b^2)$, satisfying
\be
V (X \pm b^2) \big|_{B_{\mu\nu} = b_{\mu\nu}} = 0 \,,\quad V' (X \pm b^2) \big|_{B_{\mu\nu} = b_{\mu\nu}} = 0 \,. \label{potential1} \\
\ee
Imposing $b_{\mu\nu} b^{\mu\nu} = -b^2$ gives
\be
\beta = \frac{b\sqrt{h}}{\sqrt{2 f}} \,, \label{sol1}
\ee
for which $H_{\mu\nu\rho}=0$. Substituting Eqs.~(\ref{metric})--(\ref{bfield}) and (\ref{sol1}) into the EOMs~(\ref{eom1})--(\ref{eom2}) yields the reduced equations summarized in \ref{appendixA}.
From the (\ref{ett}) and (\ref{err}) one obtains
\be
f = c_0 h \,, \label{result3}
\ee
with integration constant $c_0$. Substituting Eq.~(\ref{result3}) back and combining Eqs~(\ref{err})--(\ref{euu}), one find
\begin{align}
h &= 2 k \Big[b^2 \gamma_1 \left(b^2 \gamma_2 (n-4)-3 n+4\right)+2 \left(b^2 \gamma_2-1\right) \nn \\
&\quad \left(b^2 \gamma_2 (n-4)-n\right)\Big]\Big/\Big[c_0 \left(b^2 \gamma_1+2 b^2 \gamma_2-2\right)  \nn \\
&\quad  \left(b^2 \gamma_1 (3 n-4) +2 b^2 \gamma_2 (n-4)-2 n\right)\Big] - m r^{1-n} \,, \label{hresult}
\end{align}
where $m$ is another integration constant, interpreted as the mass parameter. Finally, substituting Eq.~\eqref{hresult} into the gravitational field equations~\eqref{ett}--\eqref{euu} and the KR field EOM~\eqref{etr} yields two independent constraints
\begin{align}
&b^4 k (n-2) (n-1) n \gamma_1 \gamma_2 =0 \,, \label{finaleom1} \\
&b^2 \gamma_1 \gamma_2 k (n-1) n \left(b^2 (\gamma_1+2 \gamma_2) (3 n -4) - 2 n\right) =0 \,. \label{finaleom2}
\end{align}
We have restricted to $b\ne0$ and $n\ge2$. Nevertheless, the above two equations still admit some special possibilities. For example, when $k=0$, corresponding to the planar horizon geometry, both equations are automatically satisfied. Another trivial possibility is $\gamma_1=\gamma_2=0$, for which the KR field is completely decoupled from the metric field and the theory reduces to GR. If, however, one is interested in nontrivial solutions in $D\ge4$ ($n\ge2$) that do not rely on a particular horizon geometry, these equations imply $\gamma_1=0$ or $\gamma_2=0$. The theory therefore separates into two allowed single-coupling sectors. Introducing $\gamma_1=2\ell_1/b^2$ and $\gamma_2=\ell_2/b^2$, these sectors are
\begin{align}
\textup{Case 1:}&\quad \ell_2 = 0 \,, \\
\textup{Case 2:}&\quad \ell_1 = 0  \,. \label{branchx} 
\end{align}
The corresponding black hole solutions are
\begin{align}
\textup{Case 1:}&\quad  h = \frac{k}{c_0 (1 - \ell_1)} - \frac{m}{r^{n-1}} \,, \quad f = c_0 h \,,\quad \beta = \frac{b}{\sqrt{2 c_0}} \,,  \label{branch1} \\
\textup{Case 2:}&\quad  h = \frac{k}{c_0 } - \frac{m}{r^{n-1}} \,, \quad f = c_0 h \,,\quad \beta = \frac{b}{\sqrt{2 c_0}} \,. \label{branch2} 
\end{align}

The first class with $n=2$ and $k=1$ was reported in Ref.~\cite{Yang:2023wtu}, where $c_0$ was set to unity. However, the first two subtleties we pointed out in the Introduction were not carefully addressed in previous works. In particular, the full KR field equations were not explicitly verified, and the authors~\cite{Yang:2023wtu} adopted the oversimplified argument in Ref.~\cite{Lessa:2019bgi}, which asserted that the $\ell_{2}R$ term could be absorbed into the Einstein-Hilbert term through a redefinition and was therefore omitted, effectively setting $\ell_{2}=0$. The resulting solution remains valid; however, as indicated by Eqs.~\eqref{finaleom1}--\eqref{finaleom2}, the $\ell_2$ cannot take nonvanishing values when $\ell_1 \ne 0$. It is precisely for these reasons---and because only the special four-dimensional case with $n=2$ was considered in the literature, where the gravitational field equations hold even without verifying the KR sector---that the second class of the solution~(\ref{branch2}) was not obtained in previous works.

In the case of $\ell_1 = 0$, the solution is nothing but a Schwarzschild metric. However, this does not imply that either the theory or the black hole solution itself degenerates into GR directly, since the parameter $m$ does not represent the physical mass of the black hole. A more detailed discussion of this class will be presented in the next section, after the conserved mass is properly defined.

Finally, by rescaling $t$ and $m$, the constant $c_0$ can be absorbed, yielding
\begin{align}
\textup{Case 1:}&\quad h = k - \frac{m}{r^{n-1}} \,, \quad f = \frac{h}{1 - \ell_1}  \,,\quad \beta = \frac{b\sqrt{1 - \ell_1}}{\sqrt{2 }} \,,  \label{branch1a} \\
\textup{Case 2:}&\quad  h = k - \frac{m}{r^{n-1}} \,, \quad f = h \,,\quad \beta = \frac{b}{\sqrt{2 }} \,. \label{branch2a} 
\end{align}

\section{Thermodynamics} \label{Thermodynamics}

We now analyze the thermodynamics of the two branches. The event horizon at $r=r_h$ is determined by $f(r_h)=0$. The Hawking temperature follows from the surface gravity associated with the Killing vector $\xi=\partial_t$,
\be
T = \frac{K}{2\pi} \,, \quad  K^2 = - \frac{ \nabla^\mu \xi^\nu \nabla_\mu \xi_\nu }{2} \Big|_{r = r_h} \,.\label{surface gravity and T}
\ee
This yields
\bea
\textup{Case 1:}&&\quad  T_1 = \frac{h'(r_h)}{4\pi \sqrt{1 - \ell_1}} = \frac{(n-1)k}{4\pi r_h \sqrt{1 - \ell_1}} \,,\\
\textup{Case 2:}&&\quad  T_2 = \frac{h'(r_h)}{4\pi} = \frac{(n-1)k}{4\pi r_h } \,,
\eea
The entropy is computed using the Wald formula,
\be
S = -\frac{\pi }{\kappa} \int_{\Omega_{n,k}} r^n \frac{\partial L}{\partial R_{\mu\nu\rho\sigma}}\epsilon_{\mu\nu}\epsilon_{\rho\sigma} \bigg|_{r\to r_h}= \frac{2 \pi w_{n,k} }{\kappa}   (1 -\ell_1 -\ell_2) r_h^n \,, \label{waldentropy}
\ee
where $\kappa = 8\pi$, the binormal $\epsilon_{\mu\nu} $ and the surface area $w_{n,1}$ for $n$-sphere are given by 
\be
\epsilon_{\mu\nu} = \sqrt{\frac{h}{f}} (\delta^t{}_\mu \delta^r{}_\nu -\delta^r{}_\mu \delta^t{}_\nu ) \,, \quad w_{n,1} = \frac{2 \pi ^{\frac{n+1}{2}}}{\Gamma \left(\frac{n+1}{2}\right)} \,.
\ee
Here, $\Gamma$ denotes the Gamma function. The resulting entropies are
\bea
\textup{Case 1:}&&\quad  S_{\textup{W}1} = \frac{2\pi w_{n,k}}{\kappa} (1 - \ell_1) r_h^n \,, \\
\textup{Case 2:}&&\quad  S_{\textup{W}2} = \frac{2\pi w_{n,k}}{\kappa} (1 - \ell_2) r_h^n \,.
\eea
The Wald formula applies provided the fields remain regular on the horizon. In EKR gravity, the KR field~(\ref{sol1}) is finite at $r=r_h$, and no additional contributions arise, in contrast to known pathologies in Horndeski or Einstein-bumblebee theories~\cite{An:2024fzf,Chen:2025ypx,Li:2025tcd,Feng:2015oea,Feng:2015wvb}. 

Since the entropy differs from the Bekenstein-Hawking area law used in Ref.~\cite{Yang:2023wtu}, the Komar mass adopted there is not expected to satisfy a consistent first law together with $T$ and $S$. Similar discrepancies have been observed in other modified gravity theories~\cite{Lu:2013ura,Chow:2013gba,Wu:2015ska,Ma:2022nwq,Liu:2022wku}. We therefore employ the Wald formalism~\cite{Wald:1993nt,Iyer:1994ys} to compute the conserved mass and establish the first law in a self-consistent manner.

\subsection{Wald formalism} \label{Wald}

The conserved charges and first law are derived using the covariant phase space (Wald) formalism~\cite{Wald:1993nt,Iyer:1994ys}, which provides a unified definition of mass and entropy in diffeomorphism-invariant theories. This approach has been extensively applied to higher-derivative gravities~\cite{Ma:2022nwq,Liu:2022wku,Fan:2014ala,Ma:2023qqj,Hu:2023gru,Ma:2024ynp,Liu:2014dva,Guo:2025muo,Guo:2025ohn,Chen:2025ary}, scalar-tensor~\cite{Liu:2013gja,Lu:2014maa,Feng:2015oea,Feng:2015wvb} and vector-tensor theories~\cite{An:2024fzf,Chen:2025ypx,Li:2025tcd,Luo:2026duf}, as well as Maxwell theory and its extensions~\cite{Gao:2003ys,Li:2016nll,Liu:2014tra,Fan:2014ala,Lu:2013ura,Chow:2013gba}.

For a Lagrangian $D$-form $\mathbf{L}(\psi)$, its variation takes the form
\be
\delta \mathbf{L} = \mathbf{E}_\psi \delta \psi + d \mathbf{\Theta} (\psi, \delta \psi) \,,
\ee
where $\mathbf{\Theta}$ is the symplectic potential. The associated Noether current $(D-1)$-form is
\begin{equation}
\mathbf{J}_\xi = \mathbf{\Theta}(\psi,\mathscr{L}_\xi \psi) - i_\xi \mathbf{L}\,,
\end{equation}
and satisfies $\mathbf{J}_\xi = d\mathbf{Q}_\xi$ on-shell. The variation of the Hamiltonian is then given by
\begin{equation}
\delta \mathcal{H} = \frac{1}{2\kappa} \int_{\Omega_{D-2}} \left( \delta \mathbf{Q}_\xi - i_\xi \mathbf{\Theta} \right)\,,
\label{WaldH}
\end{equation}
which leads to the first law from $\delta \mathcal{H}_\infty = \delta \mathcal{H}_{r_h}$.

For the EKR theory, a straightforward computation yields
\begin{equation}
\delta \mathbf{Q}_\xi - i_\xi \mathbf{\Theta} 
= -\frac{n r^{n-1}\sqrt{h}}{\sqrt{f}} (1-\ell_1-\ell_2)\, \delta f \, \Omega_{n,k}\,.
\end{equation}
Evaluating Eq.~(\ref{WaldH}) at infinity gives
\begin{align}
\textup{Case 1:}&\quad \delta M_1 = \frac{n w_{n,k}}{\kappa}\sqrt{1-\ell_1}\, \delta m \,,\\
\textup{Case 2:}&\quad \delta M_2 = \frac{n w_{n,k}}{\kappa}(1-\ell_2)\, \delta m \,,
\end{align}
from which the conserved masses follow as
\begin{align}
\textup{Case 1:}&\quad M_1 = \frac{n w_{n,k}}{2\kappa}\sqrt{1-\ell_1}\, m \,,\\
\textup{Case 2:}&\quad M_2 = \frac{n w_{n,k}}{2\kappa}(1-\ell_2)\, m \,.
\end{align}
At the horizon, one finds
\begin{equation}
\delta \mathcal{H}_{r_h} = T \delta S \,,
\end{equation}
which reproduces the Wald entropy obtained. Therefore, both branches satisfy the first law
\begin{equation}
\delta M = T \delta S \,, \label{1stlaw}
\end{equation}
together with the Smarr relation
\begin{equation}
M = \frac{n}{n-1} TS \,. \label{Smarr}
\end{equation}

Expressing the solutions in terms of the physical mass $M$, the metrics become
\begin{align}
\textup{Case 1:}\quad ds_1^2 &= -\left(k - \frac{2\kappa M_1}{n w_{n,k} r^{n-1}\sqrt{1-\ell_1}}\right) dt^2 \nn \\
&\quad+ \frac{1-\ell_1}{k - \frac{2\kappa M_1}{n w_{n,k} r^{n-1}\sqrt{1-\ell_1}}} dr^2 
+ r^2 d\Omega_{n,k}^2 \,,  \label{metric2a} \\
\textup{Case 2:}\quad ds_2^2 &= -\left(k - \frac{2\kappa M_2}{n w_{n,k} r^{n-1}(1-\ell_2)}\right) dt^2 \nn \\
&\quad+ \frac{dr^2}{k - \frac{2\kappa M_2}{n w_{n,k} r^{n-1}(1-\ell_2)}} 
+ r^2 d\Omega_{n,k}^2 \,. \label{metric2b} 
\end{align}
For $\ell_1=0$, it is straightforward to see that by introducing new time and radial coordinates
\be
\tilde{t} = t (1-\ell_2)^{\frac{1}{n-1}} \,,\quad \tilde{r} = r (1-\ell_2)^{\frac{1}{n-1}} \,,\quad 
\ee
the metric can be rewritten as
\be
d\tilde{s}_2^2 = (1-\ell_2)^{-\frac{2}{n-1}} ds_{\textup{Sch}}^2 \,.
\ee
The solution is conformally equivalent to the Schwarzschild metric, indicating equivalence to GR in vacuum after performing a conformal-like transformation on the metric, in a manner analogous to the transformation from the Jordan frame to the Einstein frame in scalar-tensor gravity~\cite{Li:2025gna}. Nevertheless, we stress that this equivalence holds only in vacuum. Once the dynamical KR field is coupled to matter---for instance, in the interior of a compact star---the mass-radius relation and other physical quantities are modified, leading to observable deviations from GR. Thus, the second class of solutions obtained in our analysis are by no means trivial.

In contrast, for $\ell_2 = 0$, the spacetime exhibits a solid angle deficit at large $r$, characteristic of a global monopole-like geometry~\cite{Barriola:1989hx}. 
For sufficiently small $\ell_1$, the deviation from GR can be weak and remain observationally viable. However, the Wald formalism analysis shows that the physical mass is given by the Noether charge rather than the Komar mass adopted in Ref.~\cite{Yang:2023wtu}. This difference implies that existing constraints on $\ell_1$ may require revision. We therefore re-evaluate the observational bounds based on the Noether mass.

\section{Implications of the Noether mass for observational constraints} \label{solar}

The metric expressed in terms of the Noether mass differs from that written using the Komar mass~\cite{Yang:2023wtu}, leading to distinct predictions for observable quantities. This provides a direct probe of how the definition of conserved charge enters observable predictions.  We therefore re-examine the corresponding observational constraints.

As a representative Solar System test, we consider the perihelion precession of Mercury in the first class of spacetime~(\ref{metric2a}). Setting $n=2$ and $k=1$, with $u=\cos\theta$, the metric reduces to
\be
ds_1^2 = - \left(1 - \frac{2 M_1}{r \sqrt{1 -\ell_1} }\right) dt^2 + \frac{1 - \ell_1}{1 - \frac{2 M_1}{r \sqrt{1 -\ell_1} }} dr^2 + r^2 (d\theta^2 + \sin^2\theta d\phi^2 ) \,.\label{d4metric}
\ee
The motion of a test particle can be derived from the standard relativistic point-particle action~\cite{Li:2020wse}, and is equivalently described by timelike geodesics in the spacetime~(\ref{d4metric}). Restricting to the equatorial plane $\theta=\pi/2$, the conserved energy and angular momentum per unit mass are
\be
{\cal E} = -\frac{P_t}{\mu} = -g_{tt} \dot{t} \,, \quad {\cal L} = \frac{P_\phi}{\mu} = g_{\phi\phi} \dot{\phi} \,.
\ee
together with the normalization condition
\be
g_{\mu\nu} \dot{X}^\mu \dot{X}^\nu = -1 \,.
\ee
Combining these relations yields the radial equation
\be
\dot{r}^2 +\frac{\left(\frac{\mathcal{L}^2}{r^2}+1\right) \left(1-\frac{2 M_1}{r \sqrt{1-\ell_1}}\right)}{1-\ell_1}-\frac{\mathcal{E}^2}{1-\ell_1} = 0 \,.
\ee
Introducing the dimensionless variable $x = \mathcal{L}^2 /(M_1 r)$, one obtains the orbital equation
\be
\frac{d^2 x}{d\phi^2} +  \frac{x}{1-\ell_1}-\frac{1}{\left(1-\ell_1\right)^{3/2}} -\frac{3 M_1^2 x^2}{\left(1-\ell_1\right)^{3/2} \mathcal{L}^2} = 0 \,.
\ee 
Treating the last term as a small correction, we expand $x = x_0 + x_1$, where
\bea
\frac{d^2 x_0}{d\phi^2} +  \frac{x_0}{1-\ell_1}-\frac{1}{\left(1-\ell_1\right)^{3/2}} &=& 0 \,, \\
\frac{d^2 x_1}{d\phi^2} +  \frac{x_1}{1-\ell_1}-\frac{3 M_1^2 x_0^2}{\left(1-\ell_1\right)^{3/2} \mathcal{L}^2}  &=& 0 \,.
\eea
The zeroth-order solution is
\be
x_0 = (1-\ell_1)^{-1/2} \left(1+ e \cos \left(\frac{\phi }{\sqrt{1-\ell _1}}\right) \right) \,,
\ee
while the first-order correction reads
\begin{align}
x_1 &=  \frac{3 M_1^2}{\left(1-\ell_1\right)^{3/2} \mathcal{L}^2} \Bigg(1+\frac{e^2}{2}-\frac{e^2 }{6} \cos \left(\frac{2 \phi }{\sqrt{1-\ell_1}}\right) \nn \\
&\quad +\frac{e \phi }{\sqrt{1-\ell_1}} \sin \left(\frac{\phi }{\sqrt{1-\ell_1}}\right)\Bigg) \,.
\end{align}
Keeping only the secularly growing term, the total solution can be recast into an effective elliptical form,
\be
x \approx  (1-\ell_1)^{-\frac12} \left[1 + e \cos \left(\frac{\phi}{\sqrt{1-\ell_1}} \left(1-\frac{\alpha}{(1-\ell_1)} \right)\right)\right] \,,
\ee
where $\alpha = 3 M_1^2/\mathcal{L}^2 $. 
It is worth noting that the total orbit, computed to first order using the Noether mass, differs from the corresponding orbit derived using the Komar mass~\cite{Yang:2023wtu}. Evidently, the period $\Phi$ of the orbit also differs from previous results, and is then given by
\be
\Phi = \frac{2 \pi \sqrt{1-\ell_1}}{ 1 - \alpha (1-\ell_1)^{-1}}  \,. 
\ee
It is worth noting that, if one directly considers the correction to the orbital period predicted by GR due to the breaking of Lorentz invariance, one can perform a Taylor expansion in $\ell_1$, as it is necessarily a small quantity, yielding
\be
\Phi = \frac{2\pi}{1-\alpha} -\frac{(1-3\alpha) \pi \ell_1}{(1-\alpha)^2} + {\cal O}(\ell_1^2) \,.
\ee
It is evident from this expression that an inappropriate definition of mass can lead to significantly different corrections to GR predictions in this theory. 

However, in the weak-field regime, $\alpha$ is already a small parameter, i.e. GR itself represents a correction to Newtonian gravity. The leading-order correction arising from the breaking of Lorentz invariance relative to Newtonian gravity is unaffected by the choice of mass definition, since the influence of the Lorentz-violating parameter on the mass definition is itself suppressed by $\alpha$. To examine this explicitly, one may further perform a Taylor expansion of the above expression in $\alpha$. Thus,
\be
\Phi = 2 \pi + 2 \pi \alpha -\ell_1 \pi  \,,
\ee
where the second term corresponds to $\Delta \Phi_{\textup{GR}}$ and the third term to $\Delta \Phi_{\ell_1}$. This result is consistent with previous findings~\cite{Yang:2023wtu}, with differences arising only at higher-order corrections.

\section{Generalization to $\Lambda\ne 0$} \label{adssolution}

\subsection{Exact solutions}

We now construct exact static black hole solutions with general topological horizons for the theory~(\ref{gravity}) in $D = n+2$ dimensions, allowing for a nonvanishing cosmological constant $\Lambda \neq 0$. The metric $g_{\mu\nu}$ and KR field $B_{\mu\nu}$ are assumed to take the same ansatz as in Eqs.~(\ref{metric})--(\ref{bfield}).

In contrast to the $\Lambda = 0$ case, obtaining exact solutions now requires relaxing the vacuum conditions~(\ref{potential1}), namely allowing $V' \neq 0$. This can be understood directly from the EOMs. As noted in the Introduction, Eqs.~(\ref{eom1})--(\ref{eom2}) can be viewed as two independent constraints on the curvature. Turning on $\Lambda$ modifies Eq.~(\ref{eom1}) while leaving Eq.~(\ref{eom2}) unchanged. Since the KR field configuration is kept the same as in the $\Lambda=0$ case, consistency of Eq.~(\ref{eom2}) requires compensating the effect of the cosmological constant through the potential. This necessitates a nonvanishing $V'$.

Following Refs.~\cite{Yang:2023wtu,Bluhm:2007bd,Maluf:2020kgf}, we adopt $V= \lambda (B_{\mu\nu} B^{\mu\nu} + b^2)$, with $V' (x) \equiv dV(x)/dx = \lambda $. The resulting solutions again fall into two branches:
\begin{align}
\textup{Case 1:}&\quad  h = k - \frac{m}{r^{n-1}} -\frac{2 \Lambda r^2}{n (n+1)} \,, \quad f = \frac{h}{1 - \ell_1}  \,,  \nn \\
 &\quad \beta = \frac{b\sqrt{1 - \ell_1}}{\sqrt{2 }} \,,\quad \lambda = \frac{4 \Lambda \ell_1}{n (1-\ell_1) b^2} \,, \label{branch1c} \\
\textup{Case 2:}&\quad  h = k - \frac{m}{r^{n-1}} -\frac{2 \Lambda r^2}{n (n+1) (1-\ell_2)} \,, \quad f = h \,, \nn \\
&\quad \beta = \frac{b}{\sqrt{2 }}\,,\quad  \lambda = \frac{2 (n+2) \Lambda \ell_2}{n (1-\ell_2) b^2}  \,. \label{branch2c} 
\end{align}
Unlike the $\Lambda=0$ case, $\ell_2$ explicitly enters the metric in Case 2, reflecting the nontrivial interplay between the KR sector and the cosmological constant. This behavior is expected, since in the absence of matter, a conformal transformation of the metric can simultaneously rescale the cosmological constant $\Lambda$ into an effective value $\Lambda_e$, thereby rendering the spacetime equivalent to GR.

\subsection{Thermodynamics}

We now present the thermodynamic properties. Treating $\Lambda$ as a fixed parameter, the temperature, entropy, and mass are
\bea
\textup{Case 1:}&& T_1 =  \frac{{k} (n-1) n  -2 \Lambda  r_{h}^2}{4 n \pi  r_{h} \sqrt{1-\ell_1} }  \,, \\
&\quad& S_1 = \frac{2\pi w_{n,k}}{\kappa} (1 - \ell_1) r_h^n \,, \\
&\quad& M_1 = \frac{n m w_{n,k} }{2\kappa} \sqrt{1 - \ell_1} \,, \\
\textup{Case 2:}&&  T_2 = \frac{{k} (n-1) n (1-\ell_2 ) -2 \Lambda  r_{h}^2}{4 n \pi  r_{h} (1-\ell_2)} \,, \\
&\quad& S_2 = \frac{2\pi w_{n,k}}{\kappa} (1 - \ell_2) r_h^n \,, \\
&\quad& M_2 = \frac{n m w_{n,k} }{2\kappa} (1 - \ell_2) \,.
\eea
The first law takes the same form as in Eq.~(\ref{1stlaw}).
In the extended phase space, identifying
\be
P = -\frac{\Lambda}{8\pi} \,,
\ee
the  thermodynamic volumes are
\be
V_1 = \frac{r_h^{n+1} \sqrt{1-\ell_1} w_n}{n+1} \,, \quad
V_2 = \frac{r_h^{n+1}  w_n}{n+1} \,. 
\ee
and the first law becomes
\be
\delta M = T \delta S + V \delta P \,, 
\ee
with the Smarr relation
\be
M = \frac{n}{n-1} T S  -\frac{2}{n-1} V P \,. 
\ee
For the case with $k = 0$, the generalized Smarr relation~\cite{Liu:2015tqa} reduces to
\be
M = \frac{n}{n+1} T S  \,.
\ee

\section{Conclusion} \label{conclusion}

We have revisited black hole solutions and their thermodynamics in EKR gravity, clarifying several subtle issues in the construction of exact solutions. Two classes of static black holes with general topological horizons were obtained in arbitrary dimensions, both with and without a cosmological constant. One branch reproduces known results in four dimensions, while the other represents a new family of solutions. Using the Wald formalism, we computed the Noether mass and showed that it differs from commonly adopted mass definitions in the literature. While such differences are negligible in the weak-field regime, they can become relevant in strong-gravity systems.  In particular, although different mass definitions lead to negligible differences in the weak-field regime, they can become important in strong-gravity systems such as black holes and neutron stars, or in observational tests dominated by relativistic effects~\cite{Will:2014kxa,Berti:2015itd}. A proper identification of the gravitational mass is therefore essential for reliably constraining Lorentz symmetry violation in such contexts. These results suggest that existing phenomenological constraints within EKR gravity, as well as other Lorentz-violating gravities, may need to be revisited when the Noether charge is taken as the physical mass. The exact solutions and their associated thermodynamic quantities presented here provide a useful starting point for further investigations of Lorentz symmetry breaking in the EKR framework.

Several directions remain to be explored. First, constructing self-consistent compact objects, such as neutron stars, is crucial~\cite{Lessa:2025kln,Luo:2026oxw}, which enables testing Lorentz-violation parameters of EKR gravity in realistic astrophysical settings. Second, since the vacuum is not asymptotically Minkowski, the stability of black holes---particularly regarding ghost instabilities or additional dynamical degrees of freedom induced by the nonminimal coupling~\cite{Hell:2021wzm,Hell:2026mle}---deserves study. Finally, nontrivial boundary effects may arise from the higher-derivative-like structure indicated by $g_{tt} g_{rr}\neq 1$, motivating extensions to include a cosmological constant and potential holographic applications.

\section*{Acknowledgements}
We are grateful to H. L\"u, and Ke Yang for useful discussions. 
S.L. is supported in part by the National Natural Science Foundation of China (No. 12105098, No. 12481540179, No. 11947216, No. 12075084, No. 11690034, and No. 12005059), and the Natural Science Foundation of Hunan Province (No. 2022JJ40264), and the innovative research group of Hunan Province under Grant No. 2024JJ1006, and by the Excellent Young Scholars Program of the Hunan Provincial Department of Education under Grant No. 25B0092. 
H.D.L.~is  supported in part by Postdoctoral Innovation Project of Shandong Province SDCX-ZG-202503036 and National Natural Science Foundation of China Grants No.12447134. 
Z.X.Y is supported in part by the Young Scholars Startup Fund of Jining Normal University.

\appendix

\section{The nonvanishing components of EOMs}
 \label{appendixA}
 The nonvanishing components of EOMs are given by
\begin{align}
E^t{}_t &\equiv h'' -\frac{\left(h'\right)^2}{2 h} +h' \left(\frac{f'}{2 f}+\frac{n \left(\gamma_1+4 \gamma_2\right) }{2 \left(\gamma_1+2 \gamma_2\right) r}\right) + \frac{n h \left(b^2 \gamma_2+1\right) f'}{b^2 \left(\gamma_1+2 \gamma_2\right) f r} \nn \\
& +\frac{n (n-1) h \left(f \left(2-b^2 \left(\gamma_1-2 \gamma_2\right)\right)-2 k \left(b^2 \gamma_2+1\right)\right)}{2 b^2 \left(\gamma_1+2 \gamma_2\right) f r^2} = 0 \,, \label{ett} \\
E^r{}_r &\equiv h'' -\frac{\left(h'\right)^2}{2 h} + h' \left(\frac{f'}{2 f} +\frac{n \left(b^2 \gamma_2+1\right)}{b^2 \left(\gamma_1+2 \gamma_2\right) r}\right) +\frac{n \left(\gamma_1+4 \gamma_2\right) h f'}{2 \left(\gamma_1+2 \gamma_2\right) f r} \nn\\
& +\frac{n (n-1) h \left(f \left(2-b^2 \left(\gamma_1-2 \gamma_2\right)\right)-2 k \left(b^2 \gamma_2+1\right)\right)}{2 b^2 \left(\gamma_1+2 \gamma_2\right) f r^2} = 0 \,,  \label{err}\\
E^u{}_u &\equiv  h'' -\frac{\left(h'\right)^2}{2 h} + h' \left(\frac{f'}{2 f}+\frac{n-1}{r}\right)+\frac{(n-1) h f'}{f r} \nn \\
& +\frac{h (n-1) (n-2)}{r^2} \left(1-\frac{2 k \left(b^2 \gamma_2-1\right)}{f \left(b^2 \left(\gamma_1+2 \gamma_2\right)-2\right)}\right)= 0 \,,  \label{euu} \\
E^{tr}_B &\equiv h''-\frac{\left(h'\right)^2}{2 h}+ h' \left(\frac{f'}{2 f}+\frac{n \left(\gamma_1+4 \gamma_2\right) }{2 \left(\gamma_1+2 \gamma_2\right) r}\right)+\frac{n \left(\gamma_1+4 \gamma_2\right) h f'}{2 \left(\gamma_1+2 \gamma_2\right) f r} \nn \\
& +\frac{ 2 n (n-1) (f-k)\gamma_2 h}{\left(\gamma_1+2 \gamma_2\right) f r^2} = 0 \,, \label{etr}
\end{align}
where, in deriving Eq.~(\ref{euu}), we have divided by the coefficient of $h''$, namely $b^2\gamma_1+2b^2\gamma_2-2$, which is assumed to be nonzero.



\end{document}